\documentclass[11pt]{article}

\usepackage[numbers,sort&compress]{natbib}
\usepackage[dvips]{color}
\usepackage{epsfig}
\usepackage{amsmath}
\usepackage{amsfonts}
\usepackage{dsfont}
\usepackage{caption}
\usepackage{graphicx}
\usepackage{dcolumn}
\usepackage{bm}
\usepackage{epstopdf}
\usepackage{float}
\usepackage{subcaption}
\usepackage{pbox}
\usepackage{array}
\date{}
\begin{document}

\title{\bf Linear-Quadratic GUP and Thermodynamic Dimensional Reduction }
\author{H. Ramezani\thanks{e-mail: h.ramezani01@umail.umz.ac.ir}\hspace{.3cm}and\hspace{.3cm}
K. Nozari\thanks{e-mail: knozari@umz.ac.ir (Corresponding Author)}
\\\\{\small {\it Department of Theoretical Physics, Faculty of Sciences,
University of Mazandaran,}}\\{\small {\it P.O. Box 47416-95447,
Babolsar, Iran}}}\maketitle

\begin{abstract}
In this paper we investigate the statistical mechanics within the Linear-Quadratic GUP (LQGUP, i.e, GUP with linear and quadratic terms in momentum) models in the semiclassical regime. Then, some thermodynamic properties of a system of 3-dimensional harmonic oscillators are investigated by calculating the deformed partition functions. According to the equipartition theorem, we show that the number of accessible microstates decreases sharply in the very high temperatures regime. When the thermal de Broglie wavelength is of the order of the Planck length, three degrees of freedom are frozen in this setup. In other words, it is observed that there is an effective reduction of the degrees of freedom from $6$ to $3$ for a system of 3D harmonic oscillators in this framework. The calculations are carried out using both approximate analytical and exact numerical methods. The results of the analytical method are also presented in the form of thermal wavelengths for better understanding. Finally, the case of a 2-dimensional harmonic is treated as another example to comprehend the results, leading to a reduction of the degrees of freedom from $4$ to $2$.\\
\begin{description}
\item[PACS numbers]
04.60.Bc
\item[Key Words]
Quantum Gravity Phenomenology, Thermodynamics, Generalized Uncertainty Principle, Dimensional Reduction
\end{description}
\end{abstract}
\section{Introduction}
Since Planck announced his solution to the black body radiation, a fundamental and very important quantity, namely the Planck constant ($\hbar\simeq {10^{-34}}\textrm{J}.\textrm{s}$) entered physics. Today, this fundamental constant plays a very crucial and consequential role in science, especially in theoretical high energy physics~\cite{hbar1, hbar2, hbar3}. For example, in quantum gravity phenomenology theories (in fact, approaches to quantum gravity proposal) such as string theory, loop quantum gravity, doubly special relativity (DSR), and also Noncommutative geometry, new quantities were introduced into physics that represent Planck's constant dependence of the structure of spacetime at high energy level regime~\cite{String1,String2,String3,String4,LQG1,LQG2,LQG3,LQG4,NCG1,NCG2,NCG3,DSR1,DSR2,DSR3,DSR4,DSR5}. One of these new quantities is the fundamental length scale called the Planck Length $l_{_{\rm Pl}}$, which is defined by combining $\hbar$ with the gravitational constant $G$ and the constant speed of light $c$\,, that is ($l_{_{\rm Pl}}= \sqrt{\hbar G/{c^3}\;}\approx 1.6\times 10^{-35}$\,m). Planck mass, that is, $M_{_{\rm Pl}}= {\sqrt{\hbar c/{G}}\approx 2.17\times 10^{-8}}$kg, is another such a quantity. Combining quantum field theory with gravity leads naturally to an effective cutoff (a minimal measurable length preferably of the order of the Planck length) in the ultraviolet regime. Indeed, the high energies used to probe small distances significantly disturb the space-time structure by their strong gravitational effects. Thus one common feature of all approaches to Quantum Gravity (QG) proposal is the existence of a minimal measurable length. The existence of a minimal measurable length modifies the Heisenberg uncertainty principle (HUP) to the so called generalized uncertainty principle (GUP) (and also extended uncertainty principle (EUP)). GUP can be understood in the context of string theory and supports the existence of a minimal length/maximal momentum (energy) (UV cutoff) \cite{QGML1,QGML2,GUPUV1,GUPUV2,GUPUV3,GUPUV4}.
 In the HUP framework, there isn't any essential limitation on the accuracy of measuring the position of test particles. Therefore, the minimal uncertainty in position measurement ($\Delta{\rm q_0}$) can be arbitrarily reduced even to a value of zero. On the other hand, there is a limitation in the GUP framework due to the minimal uncertainty in position measurement.
This mentioned phenomenological model, as a flat limit of quantum gravity proposal, represents the deformation for the density of states of the system at very high temperature regimes, which typically leads to self-exclusive measurements over microstates. This property leads to the reduction of the degrees of freedom which may reflect in essence a reduction of the dimensions of the space for the desired system~\cite{Fityo,NHG15,NR17,UV-Reduction1,UV-Reduction2}. For a recent and elegant work on dimensional reduction in quantum gravity, see~\cite{Carlip2017,Carlip2019}. There are several versions of GUP. The first one, quadratic GUP, modifies the standard Heisenberg Uncertainty Principle (HUP) by including an additional quadratic term in momentum. This suggests that gravity may exhibit different behavior at the minimal length scale than it does under general relativity~\cite{Chang:2001bm}. Another version, the linear and quadratic GUP (LQGUP ), incorporates both the maximum momentum and minimum length, includes both linear and quadratic terms for momentum, and is compatible with Doubly Special Relativity (DSR) theories~\cite{Vagenas:2019wzd,Bhatta2024,Bernaldez2023,Rahaman2023,Vagenas2021,Barman2021,Hemeda2024}.\\
In this interesting and challenging streamline, we are going to study the issue of thermodynamic dimensional reduction through the statistical mechanics of a $3$-D harmonic oscillator system within a version of the GUP model; the Linear-quadratic GUP, in the semiclassical regime. We use the deformed partition functions and study some thermodynamic properties of a system of three-dimensional harmonic oscillators. We'll see that the number of microstates decreases strongly in the high temperature regime (or equivalently at the low thermal de Broglie wavelength), and this may be effectively related to the reduction of the space dimensions at high energy regime. As an important result, a trace of the \emph{fractional} degrees of freedom (and possibly space dimensions) will be observed in this setup for a system of $3$-D harmonic oscillators.\\

In this paper, in Section 2, the statistical mechanics of the LQGUP framework is studied and the corresponding partition function is formulated. In section 3, the thermodynamics of a system of $3$D harmonic oscillators is studied analytically (approximated solutions) and also numerically (exact solutions) within the framework of LQGUP. Section 4 provides an analysis of a 2-dimensional harmonic oscillator thermodynamics in LQGUP setup. Section 5 is devoted to the summary and conclusions.

\section{Statistical Mechanics in LQGUP Phase Space}

The classical system's kinematics and dynamics on the phase space offer a framework for formulating statistical mechanics in the semiclassical regime. The Liouville volume is the crucial factor that determines the density of states, which can be utilized to derive all of a system's thermodynamic properties.

\subsection{Kinematics and Dynamics}

The generalized uncertainty relation that implies the existence of a nonzero minimal uncertainty in position measurement has the following form
\begin{equation}\label{PB0}
\Delta q \Delta p  \ge \frac{\hbar }{2}\big({1 + \alpha {({\Delta p})^2} + \alpha {\langle p \rangle }^2}+\gamma \big)\,.
\end{equation}
Inspired by doubly special relativity (DSR), there exists a limit on a test particle's momentum, which prevents it from being imprecise. Consequently, this may result in a maximum measurable momentum ~\cite{DSR3,Magueijo:2002am,Magueijo:2004vv}. Within this framework, the LQGUP that forecasts a minimum observable length and a maximum momentum can be expressed as follows, ~\cite{Vagenas:2019wzd,Das:2008kaa,Ali:2009zq,Das:2009hs,Das:2010zf,Basilakos:2010vs},

\begin{equation}\label{PB1}
\Delta q \Delta p  \ge \frac{\hbar }{2}\big(1-2 \kappa {\langle p \rangle} + 4{\kappa}^2 {\langle p^2 \rangle }+\gamma \big)\,.
\end{equation}
We set $\gamma=0$ For simplicity (\cite{GUPUV1,GUPUV2,GUPUV3,GUPUV4}). It should be mentioned that the constants $\alpha$ and $\kappa$ are related through dimensional analysis as [$\alpha$] = [$\kappa^2$]. In the given relation, $\kappa>{0}$ represents the LQGUP parameter, which supports both the minimal length and the maximal momentum, and it's defined as $\kappa={\kappa _0}/{ M_{_{\rm Pl}}c}= {\kappa _0}{l_{_{\rm Pl}}}/{ \hbar },$ that $M_{_{\rm Pl}}c^2\approx10^{19}$ GeV is the 4-dimensional fundamental scale. $\kappa_0$ is a dimensionless constant that can be determined through experimentats \cite{QGE2} and it's close to the unity, that is, $ \kappa_0\approx1$. This modified relation suggests that the uncertainty in position measurement, $\Delta q$, is always greater than $\Delta q_{min}=2\kappa\hbar=\hbar \sqrt{\alpha}$, and the uncertainty in momentum measurement, $\Delta p$, is always smaller than $\Delta p_{max}=\frac{1}{2\kappa}$~\cite{GUPUV3}.
 If $\kappa$ tends to zero ($\kappa \to {0} $), the Heisenberg uncertainty relation (HUP) is restored. Using this approach, we can derive the following algebraic structure for position and momentum operators
\begin{equation}\label{PB2}
[ \hat q_i,\hat p_j ] = i\hbar \left( \delta_{ij}-\kappa \Big(p\delta_{ij}+\frac{p_ip_j}{p} \Big)+ \kappa^2 \big(p^2\delta_{ij}+3p_ip_j \right),
\end{equation}
\begin{equation}\label{PB3}
\left[ \hat p_i,\hat p_j \right] = 0 =\left[\hat q_i,\hat q_j \right] \,.
\end{equation}

In ordinary quantum mechanics, the coordinates ${\mathfrak q}_i$ and momenta ${\mathfrak p}_i$ satisfy the Heisenberg algebra $\left[ \hat {\mathfrak q}_i,\hat {\mathfrak p}_j\right] = i\hbar {\delta _{ij}}$, whose classical counterpart is the Poisson algebra $\left\{ {\mathfrak q}_i,{\mathfrak p}_j \right\} = {\delta _{ij}}$. In the LQGUP framework, these relations are written as
\begin{equation}\label{PB4}
\left\{ {{q_i},{p_j}} \right\} =  \delta_{ij}-\kappa \Big(p\delta_{ij}+\frac{p_ip_j}{p} \Big)+ \kappa^2 \big(p^2\delta_{ij}+3p_ip_j\big),
\end{equation}
\begin{equation}\label{PB5}
\left\{ {{p_i},{p_j}} \right\} = 0 =\left\{ {{q_i},{q_j}} \right\}
\end{equation}

The Darboux theorem states that it is always possible to find variables $q_i$ and $p_i$ in terms of ${\mathfrak q}_i$ and ${\mathfrak p}_i$ such that the following relation holds
\begin{equation}\label{PB7}
\left\{ {{q_i},{p_j}} \right\} = f\left( {{q_i},{p_j}} \right)\,.
\end{equation}
Therefore, the Eq.~(\ref{PB4}) is a specific instance of the Darboux theorem and in the LQGUP framework $f\left( {q_i,p_j} \right) = i\hbar \left( \delta_{ij}-\kappa \Big(p\delta_{ij}+\frac{p_ip_j}{p} \Big)+ \kappa^2 \big(p^2\delta_{ij}+3p_ip_j \right)$. On the other hand, Eq.~(\ref{PB5}) which ensures via the Jacobi identity does not constrain our study. Therefore, we can conclude that the Eqs.~(\ref{PB4}) and (\ref{PB5}) are generally valid simultaneously in the presence of space noncommutativity.\\
The Jacobian of transformation for 2D-dimensional LQGUP phase space is as follows \cite{Vagenas:2019wzd}.
\begin{equation}\label{PB10}
J = \frac{{\partial ({q},{p})}}{{\partial ({{\mathfrak q}},{{\mathfrak p}})}} = \frac{d^D{q}d^D{p}}{(2\pi)^D\Big[1-\kappa p+\Big(\frac{2}{D+1}+\frac{1}{2}\Big)\kappa^2p^2\Big]^{D+1}}
\end{equation}

with the Jacobian of the transformation (\ref{PB10}) in hand, one is led to the LQGUP Liouville volume as
\begin{equation}\label{PB12}
d^3\omega=\frac{d^3q d^3p}{\left(1-\kappa p+{\kappa}^2 p^2\right)^4},\,\,\,d^{3N}\omega=\frac{d^{3N}qd^{3N}p}{\left(1-\kappa p+{\kappa}^2 p^2\right)^{4N}},
\end{equation}
Where $d^3\omega_c = d^3q\,d^3p$ represents the canonical Liouville volume, we assume that the system consisting of $N$ particles is kinematically separated.

\subsection{Number of Microstates}

In the LQGUP phase space, the microstate count is affected by the modified commutation relations between position and momentum operators. These deformations introduce noncommutativity in the phase space, leading to modifications in the distribution of microstates and consequently impacting thermodynamic properties. Understanding this qualitative aspect is useful before delving into the quantitative analysis provided by the partition function.

The number of microstates for a phase space containing a particle is given as follows\footnote{Throughout this article, we work in units $\hbar=k_{_B}=c=1$, where $k_{_B}$ and $c$ are the Boltzmann constant and speed of light in vacuum, respectively. Also, the figures are plotted in units of $ \kappa= 0.004$ and the mass is taken to be $m=100$ in appropriate units.}.
\begin{equation}\label{Nms}
\mathcal{N}_m=\frac{1}{(2\pi)^3}\int d^3\omega\,,
\end{equation}

\begin{figure}
\centering{\includegraphics[width=3.3in]{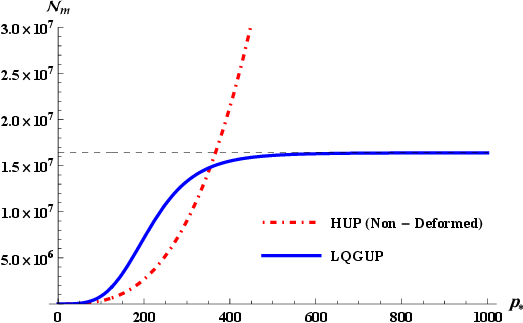}}
\hspace{3cm}\caption{\label{fig:0} The number of microstates vs. momentum $p_\ast$ for a single-particle phase space are illustrated. The red dot-dashed and blue solid lines represent microstate numbers in the non-deformed and LQGUP-deformed phase spaces, respectively. These curves coincide at low energy regime $\kappa{p_\ast}\ll{1}$ ($p_\ast\ll{
p_{_{\rm Pl}}}$) and diverge at high energy regime $\kappa{p_\ast}\sim{1}$ ($p_\ast\sim{p_{_{\rm Pl}}}$). In the LQGUP-deformed scenario, microstates behave like a zero-dimensional phase space, signifying a freezing of three degrees of freedom at high energy due to quantum gravity effects. This suggests a reduction of the degrees of freedom at high energy. The figure is plotted in units of $\kappa=\frac{\kappa_0}{p_{_{\rm Pl}}
}=0.004$ with $\kappa_0=1$ and $p_{_{\rm Pl}}=250$.}
\end{figure}

We consider the domain where momentum is restricted, i.e. where $p\leq{p_{\ast}}$ (with $p=\sqrt{\delta^{ij}p_ip_j}$ and $p_\ast$ being constant). In this scenario, it is reasonable to assume that the number of microstates is finite. As per equation (\ref{Nms}), the number of microstates for a non-deformed phase space is $\mathcal{N}_m=({{V}}/{2\pi^2})\int_0^{p_\ast}
p^2dp=({{V}}/{6\pi^2})\,p_\ast^3$, where $V$ represents the spatial volume of the system.
In the context of the LQGUP phase space, when substituting the Liouville volume (\ref{PB12}) into equation (\ref{Nms}), it is formulated as
\begin{equation}\label{LQNMS}
\begin{gathered}
\mathcal{N}_m=\frac{{V}}{2\pi^2}\int_0^{p_\ast}\frac{p^2dp}{
\big(1-\kappa p+\kappa^2p^2\big)^4}=\frac{{V}}{2\pi^2\kappa^3}
\Bigg[\frac{8\pi\sqrt{3}+81}{243}\\+{\frac{\left(\kappa p_\ast-1\right)\Big[2\kappa p_\ast \left(2\kappa p_\ast-1\right)\left(2\kappa p_\ast\left(\kappa p_\ast-1\right)+5\right)+9\Big]}{27\Big[\kappa p_\ast \left(\kappa p_\ast-1\right)+1\Big]^3}+\frac{16 \sqrt{3}}{81} \tan ^{-1}\bigg(\frac{2 \kappa p_\ast-1}{\sqrt{3}}\bigg)}\Bigg]\,.
\end{gathered}
\end{equation}
In the low temperature regime, i.e., $\kappa{p_\ast}\ll{1}$ ($p_\ast\ll{p_{_{\rm Pl}}}$), equation (\ref{LQNMS}) behaves as $\mathcal{N}_m\sim\,p_\ast^3$ for the number of microstates. This aligns with the standard result of no deformation in this regime. In the high temperature regime $\kappa{p_\ast}\sim{1}$ ($p_\ast\sim{p_{_{\rm Pl}}}$), the number of microstates (\ref{LQNMS}) is independent of $p_\ast$ and behaves as $\mathcal{N}_m\sim{cte}$ indicating a three-order reduction. This is illustrated in Fig. \ref{fig:0}, where the number of microstates (\ref{LQNMS}) is plotted versus $p_\ast$ and compared to the standard state without deformation. This result demonstrates that in the high-energy regime where quantum gravity effects (minimum length and maximal momentum) prevail, the number of microstates will significantly decrease.
Even though this outcome is qualitatively derived for a generic system (without specifying a Hamiltonian function and ensemble density) in this subsection, we will clarify this outcome explicitly in the following section for the specific scenario of a 3D-harmonic oscillator in canonical ensemble based on the renowned equipartition theorem of energy.

\subsection{Partition Function}
The Hamiltonian equations for each of the phase space coordinates $\left( q_i,p_j \right)$ are $\dot {q}_i = \left\{ q_i,H \right\}$ and $\dot {p}_i = \left\{ p_i,H \right\}$, where  $H\left( q_i,p_j\right)$ is the Hamiltonian and determines the dynamics of the whole system. Hence, for a system such as a three-dimensional harmonic oscillator, the partition function for the LQGUP framework is given by
\begin{eqnarray}\label{PB13}
\begin{gathered}
Z_N(T,V,\kappa ) = \frac{1}{h^{3N}} \int\int{\frac{{d^{3N}q\,\,{d^{3N}}p}}{{\left({1-\kappa p+{\kappa}^2 p^2}\right)}^{4N}}\exp \left( {-\frac{{H(q,p)}}{T}}\right)}\,,
\end{gathered}
\end{eqnarray}
in which the Gibbs factor $\frac{1}{N!}$ is omitted, as it is assumed that the oscillators are localized. As evident from the relation (\ref{PB13}), the fundamental volume element of the phase space, $(h\left(1-\kappa p+{\kappa}^2 p^2\right))^N$, is larger than the usual fundamental volume element $h^N$ for large momentum, where $h$ is the Planck constant. Now, let's assume that the particles in the system are also dynamically decoupled, represented by $H_N=\sum_{i} H(q_i,p^i)$, where $q_i$ and $p_i$ denote the position and momentum of the $i$-th particle. In this case, the total partition function $\mathcal{Z}_N$ for this type of system is also decoupled as $\mathcal{Z}_N = \mathcal{Z}_1^N$, where

\begin{eqnarray}\label{PB14}
\begin{gathered}
Z_1(T,V,\kappa ) = \frac{1}{h^3}\int_\Gamma d^3\omega\exp[-H/T]= \frac{1}{{{h^3}}}\int {{d^3}q \int{\frac{{{d^3}p}}
{{\left({1-\kappa p+{\kappa}^2 p^2} \right)}^4}\exp \left( { - \frac{{H(q,p)}}{T}} \right)} },
\end{gathered}
\end{eqnarray}
where $\Gamma$ is the total volume of phase space.

\section{Thermodynamics of a System of 3D Harmonic Oscillators in LQGUP Phase Space}

In this section, we study the thermodynamics of a system of 3-dimensional harmonic oscillators in the LQGUP framework, which accounts for minimal length and maximal momentum effects. The Hamiltonian function for a single harmonic oscillator
is then $H=\frac{p^2}{2m}+\frac{1}{2}m\omega^2q^2$, and by substituting it in Eq. (\ref{PB14}), the corresponding partition function for a single particle in semi-classical regime can be derived as follows

\begin{equation}\label{ho0}
\begin{gathered}
 {\mathcal Z}_1=\frac{16\pi^2}{h^3}\int_{q_{min}}^{\infty}\int_{0}^{p_{_{max}}}
\frac{\exp\big[-\frac{p^2}{2mT}\big]\exp\big[-\frac{m{w^2}q^2}{2T}\big]q^2p^2dpdq}{\big(1-\kappa p +\kappa^2 p^2\big)^4}.
\end{gathered}
\end{equation}
The partition function obtained above cannot be solved analytically. Therefore, we must rely on numerical calculation methods to determine this function and other related thermodynamic variables. However, we can approximate the answer by expanding the deformed phase space volume as follows
\begin{equation}\label{ho0.5}
  \frac{1}{\big(1-\kappa p +\kappa^2 p^2\big)^4} = 1+4\kappa p +6\kappa^2 p^2 +O\left(\kappa ^3\right),
\end{equation}
Thus, the partition function (\ref{ho0}) can be approximated as follows
\begin{equation}\label{ho1}
\begin{gathered}
 {\mathcal Z}_1\simeq \frac{16\pi^2}{h^3}\int_{0}^{\infty}\int_{0}^{\frac{1}{2\kappa}}
\exp\big[-\frac{p^2}{2mT}\big]\exp\big[-\frac{m{w^2}q^2}{2T}\big]q^2p^2 \big(1+4\kappa p +6\kappa^2 p^2\big) dp dq
\end{gathered}
\end{equation}
After integration of the above relation, we'll have
\begin{equation}\label{ho1.5}
\begin{gathered}
{\mathcal Z}_1[T]=\Big(\frac{T}{w}\Big)^3\Bigg[\sqrt{\frac{128m T\kappa^2}{\pi}}-\frac{1}{\sqrt{8\pi m T\kappa^2}}\Big(9+68 m T\kappa^2 \Big)e^{\frac{-1}{8m T\kappa^2 }}\\+\Big(1+18 m T \kappa ^2\Big) \text{erf}\Big(\frac{1}{ \sqrt{8m T\kappa^2}}\Big)\Bigg]\,.\\
\end{gathered}
\end{equation}
In this relation, ``$\text{erf}$`` is the error function, defined by: $\text{erf}(x)=(2/\sqrt{\pi})\int_{0}^{x}e^{-t^2}dt$. To better understand the qualitative features of the partition function (\ref{ho1.5}), it is convenient to rewrite it in terms of the thermal de Broglie wavelength $\lambda$, defined by: $\lambda=\frac{h}{\sqrt{2\pi{m}T}}$ as
\begin{equation}\label{ho2}
\begin{gathered}
 {\mathcal Z}_1[\lambda]\approx \Big(\frac{2\pi \ell^2}{\lambda^2}\Big)^3\bigg[\frac{4\lambda_{_{\rm Pl}}}{\sqrt{\pi}\lambda}-\frac{\lambda}{\sqrt{\pi}\lambda_{_{\rm Pl}}}\Big(9+\frac{17{\lambda_{_{\rm Pl}}}^2}{2\lambda^2}\Big)\exp \Big[\frac{-\lambda^2}{\lambda_{_{\rm Pl}}^2}\Big]+\Big(1+\frac{9{\lambda_{_{\rm Pl}}}^2}{4\lambda^2}\Big)\text{erf}\Big[\frac{\lambda}{\lambda_{_{\rm Pl}}}\Big] \bigg].
 \end{gathered}
\end{equation}
Where $\ell=\sqrt\frac{1}{m\omega}$ is the characteristic length of the harmonic oscillator system. In this formula, ${\lambda_{_{\rm Pl}}}$ is the thermal de Broglie wavelength in the LQGUP framework, which is of the order of the Planck scale thermal de Broglie wavelength,
\begin{equation}\label{lambda-GUP}
\lambda_{_{\rm Pl}}=4\sqrt {\pi}\kappa=4\sqrt {\pi}{\kappa_0}{{l_{_{\rm pl}}}}=(2\kappa_0)\frac{h}{\sqrt{
2\pi{m_{_{\rm Pl}}}T_{_{\rm Pl}}}}\,.
\end{equation}
By expanding the partition function (\ref{ho2}) for very high and low temperature regimes, we find
\begin{equation}\label{ho3}
{\mathcal Z}_1[\lambda]=\left\{
  \begin{array}{ll}
    \begin{gathered}(8{\pi}^3)\frac{{\ell}^6}{\lambda^6}=\Big(\frac{T}{\hbar \omega}\Big)^3\end{gathered} & \hspace{.5cm}\lambda
    \gg\lambda_{_{\rm Pl}},\\\\
    \begin{gathered}\Big(\frac{544\pi^{5/2}}{15}\Big)\frac{\ell^6}{\lambda^3\lambda_{_{\rm Pl}}^3}\end{gathered} &
    \hspace{.5cm}\lambda\sim\lambda_{_{\rm Pl}}\,,
  \end{array}
\right.
\end{equation}
where $\mathcal{Z}_0=\big(\frac{T}{\hbar \omega}\big)^3$ is the non-deformed partition function for a three
dimensional single-harmonic oscillator. As observed from the above relation, the influence of quantum gravity (minimal length and maximal momentum) can be disregarded at the low temperature limit $\lambda\gg\lambda_{_{\rm Pl}}\,(T\ll{T}_{_{\rm Pl}})$, and the classical outcome for the harmonic oscillator partition function is restored. Additionally, it is evident from Eq.(\ref{ho3}) that three degrees of freedom will freeze ($\lambda^6\rightarrow\lambda^3$) at the Planck length scale $\lambda\sim\lambda_{_{\rm Pl}}$ in the LQGUP framework, and the partition function behaves as if it were effectively a partition function of a $1.5$D harmonic oscillator. Therefore, it is possible to conclude an effective reduction in the degrees of freedom from $6$ to $3$ and therefore in the space dimensions from $3$ to $1.5$ in the GUP framework for harmonic oscillator systems. This feature evidently signals the existence of {\textit{fractional}} degrees of freedom and possibly space dimensions for such a system ~\cite{Frac1,Frac2}.\\
 It is important to note that in these processes of decreasing number of degrees of freedom, at intermediate temperatures in all the figures presented in this paper, we observe a fractional nature of the number of the degrees of freedom. In fact, it would be more accurate to state that the number of the degrees of freedom of the deformed models mentioned are not completely integers at high temperatures; rather, they are indeed fractional (see also figure \ref{fig:2}).\\

From Eq.(\ref{ho2}), the total partition function of the system of harmonic oscillators in the LQGUP framework can be expressed as follows\\
\begin{equation}\label{ho3.5}
\begin{gathered}
  {\mathcal Z}_N[\lambda]\approx \Big(\frac{2\pi \ell^2}{\lambda^2}\Big)^{3N}\bigg[\frac{4\lambda_{_{\rm Pl}}}{\sqrt{\pi}\lambda}-\frac{\lambda}{\sqrt{\pi}\lambda_{_{\rm Pl}}}\Big(9+\frac{17{\lambda_{_{\rm Pl}}}^2}{2\lambda^2}\Big)\exp \Big[\frac{-\lambda^2}{\lambda_{_{\rm Pl}}^2}\Big]+\Big(1+\frac{9{\lambda_{_{\rm Pl}}}^2}{4\lambda^2}\Big)\text{erf}\Big[\frac{\lambda}{\lambda_{_{\rm Pl}}}\Big] \bigg]^N.
 \end{gathered}
\end{equation}
Now, utilizing Eq. $(\ref{ho3.5})$, we can derive all thermodynamic quantities using the standard definitions. First, we begin with the Helmholtz free energy $F$, which is expressed as follows\\
\begin{equation}\label{ho4}
\begin{gathered}
F=-NT\ln \Bigg[\Big(\frac{2\pi \ell^2}{\lambda^2}\Big)^{3}\bigg[\frac{4\lambda_{_{\rm Pl}}}{\sqrt{\pi}\lambda}-\frac{\lambda}{\sqrt{\pi}\lambda_{_{\rm Pl}}}\Big(9+\frac{17{\lambda_{_{\rm Pl}}}^2}{2\lambda^2}\Big)\exp \Big[\frac{-\lambda^2}{\lambda_{_{\rm Pl}}^2}\Big]+\Big(1+\frac{9{\lambda_{_{\rm Pl}}}^2}{4\lambda^2}\Big)\text{erf}\Big[\frac{\lambda}{\lambda_{_{\rm Pl}}}\Big] \bigg]\Bigg].
 \end{gathered}
\end{equation}
\\
Having equipped with Helmholtz free energy, we can obtain other thermodynamic quantities as follows.

\subsection{Internal energy}

With the modified Helmholtz free energy as given by Eq.~(\ref{ho4}), the internal energy of the
system $U=-T^2\Big(\frac{\partial}{\partial T}\big(
\frac{F}{T}\big)\Big)_{N,V}$can be modified as follows
 \begin{eqnarray}\label{ho5}
U=NT\left(
\frac{\begin{gathered}-9+16\frac{{\lambda_{_{\rm Pl}}}^2}{\lambda^2}\bigg[\frac{{\lambda_{_{\rm Pl}}}^2} {8\lambda^2}\Big(7\exp \Big[\frac{\lambda^2} {\lambda_ {_{\rm Pl}}^2}\Big]-16\Big)-2\bigg]+3 \sqrt{\pi}\frac{{\lambda_ {_{\rm Pl}}}^3} {\lambda^3} (1+\frac{{3\lambda_ {_{\rm Pl}}}^2} {\lambda^2})\text {erf}\Big[\frac {\lambda} {\lambda_{_{\rm Pl}}}\Big]\exp \Big[\frac{\lambda^2} {\lambda_ {_{\rm Pl}}^2}\Big]\end{gathered}}
{\begin{gathered}\frac{{\lambda_{_{\rm Pl}}}^2}{\lambda^2}\bigg[-9-\Big(\frac{{17\lambda_{_{\rm Pl}}}^2}{2\lambda^2}\Big)+\frac{\sqrt{2}\lambda_{_{\rm Pl}}}{2\lambda}\exp \Big[\frac{\lambda^2} {\lambda_{_{\rm Pl}}^2}\Big]\Big(\frac{4\sqrt {2}\lambda_{_{\rm Pl}}}{\lambda}+\sqrt{2\pi}\big(1+\frac{9{\lambda_{_{\rm Pl}}}^2}{4\lambda^2}\big)\text {erf}\Big[\frac{\lambda}{\lambda_{_{\rm Pl}}}\Big]\Big)\bigg]\end{gathered}}\right).
\end{eqnarray}

\begin{figure}[ht]
  \centering
  \begin{subfigure}[b]{0.35\textwidth}
    \includegraphics[width=3.3in]{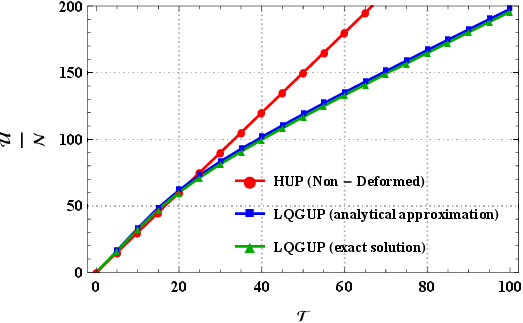}
    \caption{}
    \label{fig:sub1}
  \end{subfigure}
  \hfill
  \begin{subfigure}[b]{0.48\textwidth}
    \includegraphics[width=3.3in]{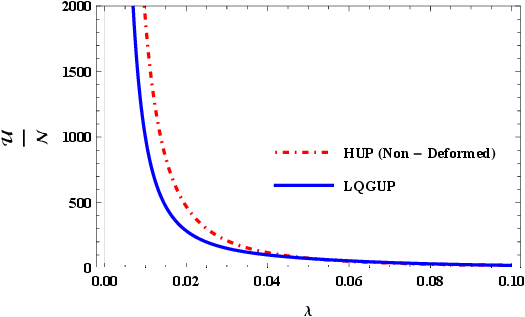}
    \caption{}
    \label{fig:sub2}
  \end{subfigure}
  \caption{ The internal energy of a $3$-D harmonic oscillator is depicted in Figure (a) as a function of temperature. It is observed that at high temperatures (Planck temperature), the system's internal energy deviates significantly from its classical state. However, as the temperature decreases, this deviation gradually diminishes until it eventually aligns with the classical state. Also, it can be observed that the approximate solution and the exact numerical solution are very close together. Figure (b) displays the internal energy in terms of thermal wavelength. It is evident that at extremely small wavelengths (minimal length effects), the deviation from the classical state intensifies. Conversely, as the thermal wavelength increases, the behavior of the internal energy approaches its classical state. Figures are plotted in units of $N=5, m=100, \omega=10, \kappa = 0.004$.}
  \label{fig:1}
\end{figure}

\begin{figure}[ht]
  \centering
  \begin{subfigure}[b]{0.35\textwidth}
    \includegraphics[width=3.3in]{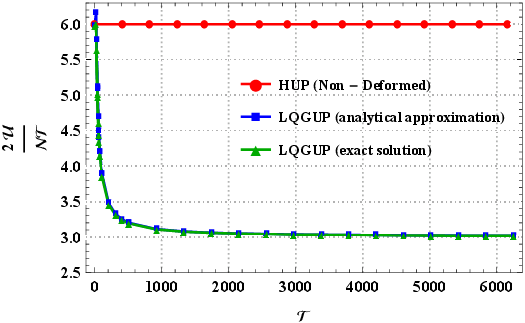}
    \caption{}
    \label{fig:sub1}
  \end{subfigure}
  \hfill
  \begin{subfigure}[b]{0.48\textwidth}
    \includegraphics[width=3.3in]{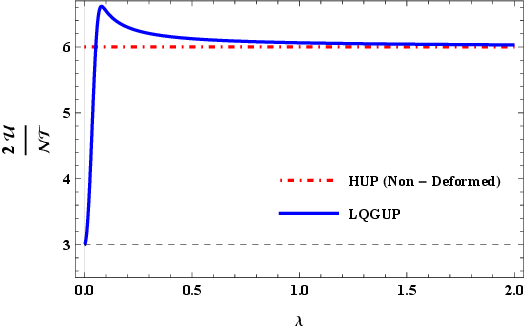}
    \caption{}
    \label{fig:sub2}
  \end{subfigure}
  \caption{The number of degrees of freedom of a 3D harmonic oscillator is plotted in Figure (a) as a function of temperature. According to the classical equipartition theorem of energy, the number of degrees of freedom of a system can be determined by the formula $\frac{(U/N)}{(T/2)}$. It is observed that in the GUP model, this value is temperature-dependent and decreases from $6$ to $3$ at extremely high temperatures for a 3D harmonic oscillator system. Figure (b) shows the number of degrees of freedom in terms of thermal wavelength. It is evident that at extremely small wavelengths, the system's three degrees of freedom are reduced.}
  \label{fig:2}
\end{figure}

For a standard 3D-harmonic oscillator, the internal energy linearly depends on the temperature through the well-known relation $U_0=3NT$. Any high energy scale is accessible for the system just by a sufficiently large increment in temperature. In figure (\ref{fig:1}), the internal energy diagram versus temperature (approximated and exact solutions) and also versus thermal de Broglie wavelength in the LQGUP framework are shown. As can be seen from the figure, as the temperature increases, the deviation from the classical state (quantum gravity effects that appear as a minimal length and maximal momentum) becomes more and more appreciable. To expand the relation (\ref{ho5}) for both of the high and low temperature regimes, we have.
\begin{equation}\label{ho6}
U=
\left\{
  \begin{array}{ll}
    3\,NT & \hspace{.5cm}\lambda
    \gg\lambda_{_{\rm Pl}},\\\\
   \frac{3}{2}\,NT &
    \hspace{.5cm}\lambda\sim\lambda_{_{\rm Pl}}\,.
  \end{array}
\right.
\end{equation}
As the above relations clearly show, the system has a reduction of the degrees of freedom in the LQGUP phase space model.
In fact, the degrees of freedom of the system, i.e. $2U/NT$ are here reduced from $6$ to $3$. The degrees of freedom are plotted in figure (\ref{fig:2}). As one can see in this figure, the reduction of the degrees of freedom disappears at lower temperatures for systems with $\lambda$ greater than the fixed $\lambda_{_{\rm Pl}}$. As we have mentioned earlier, when $\lambda$ is of the order of the $\lambda_{_{\rm Pl}}$, the system behaves exactly like a $1.5$D harmonic oscillator or equivalently with $3$ degrees of freedom.
So, as we have seen from the figures, there are fractal dimensions for intermediate temperatures, especially the fractal dimension of $1.5$ for a system of harmonic oscillators in the LQGUP because three degrees of freedom are frozen for this model.

\subsection{Specific heat}

The high temperature behavior of the internal energy in the LQGUP framework could be also more precisely understood from another important thermodynamical quantity named as specific heat, that is defined from the relation (\ref{ho5}) by the
standard definition $C_{_V}=\big(\frac{\partial U}{\partial T}
\big)_{\mbox {v}}$ as follows

\begin{equation}\label{ho7}
C_{_V}=
\frac{N\,\left({\begin{gathered}
\frac{{\sqrt{2}\lambda_{_{\rm Pl}}}}{\lambda}\Bigg[-9 +\exp\Big[\frac {\lambda^2}{\lambda_{_{\rm Pl}}^2}\Big]
\frac{56{\lambda_{_{\rm Pl}}}^6} {\lambda^6}+\frac{{\lambda_{_{\rm Pl}}}^2} {4\lambda^2}\Big(745+\frac{{2\lambda_{_{\rm Pl}}}^2}{\lambda^2}\big(1157+\frac {{553\lambda_{_{\rm Pl}}}^2}{\lambda^2}\big)\Big) \\-4\exp\Big[\frac{\lambda^2}{\lambda_{_{\rm Pl}}^2}\Big]\bigg(9+\frac{{\lambda_{_{\rm Pl}}}^2}{2\lambda^2}\Big(-17
+\frac{{\lambda_{_{\rm Pl}}}^2}{2\lambda^2}\big(259+\frac{{553\lambda_{_{\rm Pl}}}^2} {2\lambda^2}\big)\Big)\bigg)\Bigg]\\+\exp\Big[\frac{\lambda^2}{\lambda_{_{\rm Pl}}^2}\Big]\text{erf}\Big[\frac{\lambda}{\lambda_{_{\rm Pl}}}\Big]\Biggr[-\sqrt{2\pi} \Bigg[9+\frac{{\lambda_{_{\rm Pl}}}^2}{4\lambda^2}\bigg(83+\frac{{9\lambda_{_{\rm Pl}}}^2} {2\lambda^2}\Big(21+\frac{{2\lambda_{_{\rm Pl}}}^2}{\lambda^2}\big(99+\frac{{67\lambda_{_{\rm Pl}}}^2}{\lambda^2}\big)\Big)\bigg)\Bigg]
\\+\exp\Big[\frac{\lambda^2}{\lambda_{_{\rm Pl}}^2}\Big]\frac{6{\lambda_{_{\rm Pl}}}^5}{\sqrt{2}\lambda^5}\bigg(\frac {3\sqrt{\pi}{\lambda_{_{\rm Pl}}}}{\lambda}\Big(3+\frac{{31\lambda_{_{\rm Pl}}}^2}{4\lambda^2}\Big)+\pi\Big(1+\frac {{3\lambda_{_{\rm Pl}}}^2}{2\lambda^2}\Big)\Big(1+\frac{{9\lambda_{_{\rm Pl}}}^2} {2\lambda^2}\Big)\text {erf}\Big[\frac{\lambda}{\lambda_{_{\rm Pl}}}\Big]\bigg)\Biggr]
\end{gathered}}\right)}
{\begin{gathered}\frac{\sqrt {2}{\lambda_{_{\rm Pl}}}^3} {\lambda^3}\Bigg(9+\frac{{17\lambda_{_{\rm Pl}}}^2}{2\lambda^2}-\frac{\sqrt{2}\lambda_{_{\rm Pl}}}{2\lambda}\bigg(\frac{4\sqrt {2}\lambda_{_{\rm Pl}}}{\lambda}+\sqrt{2\pi}\Big(1+\frac{{9\lambda_{_{\rm Pl}}}^2} {4\lambda^2}\Big)\text{erf}\Big[\frac{\lambda}{\lambda_{_{\rm Pl}}}\Big]\exp\Big[\frac{\lambda^2}{\lambda_{_{\rm Pl}}^2}\Big]\bigg)\Bigg)^2\end{gathered}}\,.
\end{equation}

\begin{figure}[ht]
  \centering
  \begin{subfigure}[b]{0.35\textwidth}
    \includegraphics[width=3.3in]{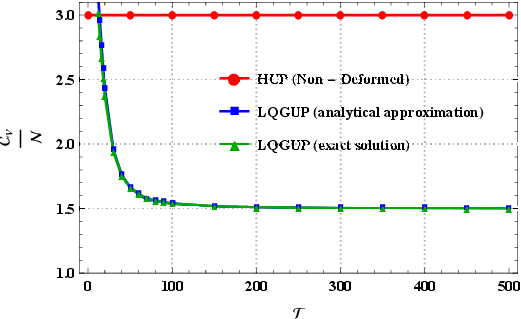}
    \caption{}
    \label{fig:sub1}
  \end{subfigure}
  \hfill
  \begin{subfigure}[b]{0.48\textwidth}
    \includegraphics[width=3.3in]{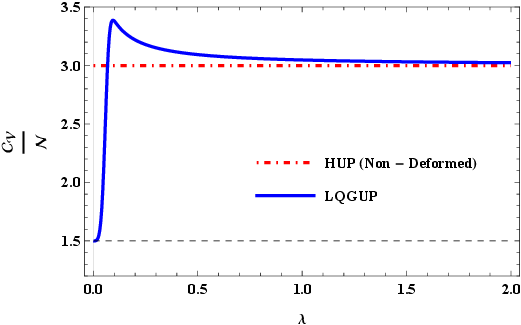}
    \caption{}
    \label{fig:sub2}
  \end{subfigure}
  \caption{The specific heat of a 3D harmonic oscillator is shown in Figure (a) as a function of temperature. The system's specific heat is temperature-dependent for the LQGUP model and asymptotically approaches $1.5$ at the very high temperature regime, which signals the effective reduction of the degrees of freedom from $6$ to $3$ in this setup. It is also clear from the figure that the specific heat is bounded as $1.5 \leq\frac{C_{_V}}{N}\leq 3$ in the LQGUP framework. also, as can be seen in this diagram, the values of the numerical and analytical solutions are very close together. Figure (b) displays the specific heat in terms of thermal wavelength. It is observed that at very small wavelengths, the deviation from the classical state intensifies and finally reaches the value of $3/2$.}
  \label{fig:3}
\end{figure}

The temperature-dependent behavior of this quantity is
plotted in figure (\ref{fig:3}). Expansion of the specific
heat (\ref{ho7}) for the low and high temperatures
regimes gives
\begin{equation}\label{ho8}
C_{_V}=
\left\{
  \begin{array}{ll}
    3\,N & \hspace{.5cm}\lambda
    \gg\lambda_{_{\rm Pl}},\\\\
   \frac{3}{2}\,N &
    \hspace{.5cm}\lambda\sim\lambda_{_{\rm Pl}},
  \end{array}
\right.
\end{equation}
which shows that the specific heat approaches $C_{_V}\rightarrow3N/2$ at the high temperature limit (see also figure (\ref{fig:3})). As it can be observed, the equation (\ref{ho7}) is a lengthy equation that gives the precise value of $\frac{3}{2}$ under a limiting operation for extremely high temperatures or equivalently for very short wavelengths, and must be taken into account to obtain the accurate value. If, for example, the Planck wavelength is neglected from the fourth order onwards, an incorrect and highly inaccurate value $\frac{26}{7}$ will be result.\\ According to {\it the classical equipartition theorem}, each of the degrees of freedom of the system has a contribution of $\frac{1}{2}\,T$ to its internal energy and $\frac{1}{2}$ to its specific heat too. As it is clear from figure (\ref{fig:3}), the specific heat is bounded as $\frac{3}{2}\leq\frac{C_{_V}}{N}\leq{3}$ in this setup.\\

As we have stated previously, when $\lambda$ is of the order of $\lambda_{_{\rm Pl}}$, the reduction of the degrees of freedom will not be temperature dependent, and the system behaves completely like a dimension of $1.5$ of harmonic oscillators, or equivalently, with $3$ degrees of freedom.
So, as we have seen from the figures, there are fractal dimensions for intermediate temperatures, especially the fractal dimension of $1.5$ for a system of harmonic oscillators in the LQGUP model because three degrees of freedom are frozen for this model.

\subsection{Entropy}

From Eq.~(\ref{ho4}), the entropy of the system $S=-\left(\frac{\partial F}{\partial T} \right)_{N,V}$ in LQGUP framework will be as follows

\begin{equation}\label{ho9}
S=\frac{N\left[\begin{gathered}
-9+\frac{16{\lambda_{_{\rm Pl}}}^2}{\lambda^2}\bigg(-2+\frac{{\lambda_{_{\rm Pl}}}^2} {8\lambda^2}\Big(-16 +7 \exp\Big[\frac {\lambda^2}{\lambda_{_{\rm Pl}}^2}\Big]\Big)\bigg)+\frac{{\lambda_{_{\rm Pl}}}^2}{\lambda^2}\Bigg(-\sqrt{\pi}\frac{{\lambda_{_{\rm Pl}}}} {2\lambda^2}\exp\Big[\frac {\lambda^2}{\lambda_{_{\rm Pl}}^2}\Big] \text {erf}\Big[\frac{\lambda}{\lambda_{_{\rm Pl}}}\Big]\bigg(-6+\ln[2\pi]\\
+\frac {9{\lambda_{_{\rm Pl}}}^2}{4\lambda^2}\Big(-8+\ln[2\pi]\Big)-2\Big(1+ \frac{{9\lambda_{_{\rm Pl}}}^2}{4\lambda^2}\Big)\ln[\sqrt{2\pi}{\mathcal Z}_1]\bigg)+\bigg(-9+\Big(-17+\frac{4{\lambda_{_{\rm Pl}}}^2}{\lambda^2}\exp\Big[\frac{\lambda^2}{\lambda_{_{\rm Pl}}^2}\Big]\Big)\bigg)\ln[{\mathcal Z}_1]\Bigg)
\end{gathered}\right]}
{\begin{gathered}\frac{{\lambda_{_{\rm Pl}}}^2}{\lambda^2}\Bigg[-9-\frac{{17\lambda_{_{\rm Pl}}}^2}{2\lambda^2}+\frac{\sqrt{2}\lambda_{_{\rm Pl}}}{2\lambda}\exp\Big[\frac{\lambda^2}{\lambda_{_{\rm Pl}}^2}\Big] \bigg(\frac{{4\sqrt{2}\lambda_{_{\rm Pl}}}}{\lambda}+\sqrt{2 \pi}\Big(1+\frac{9{\lambda_{_{\rm Pl}}}^2}{4\lambda^2}\Big)\text {erf}\Big[\frac{\lambda}{\lambda_{_{\rm Pl}}}\Big]\bigg)\Bigg]\end{gathered}}.
\end{equation}

\begin{figure}[ht]
  \centering
  \begin{subfigure}[b]{0.35\textwidth}
    \includegraphics[width=3.3in]{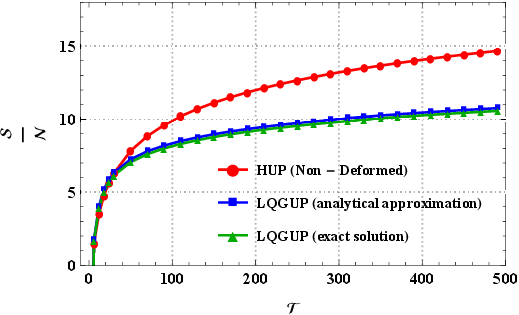}
    \caption{}
    \label{fig:sub1}
  \end{subfigure}
  \hfill
  \begin{subfigure}[b]{0.48\textwidth}
    \includegraphics[width=3.3in]{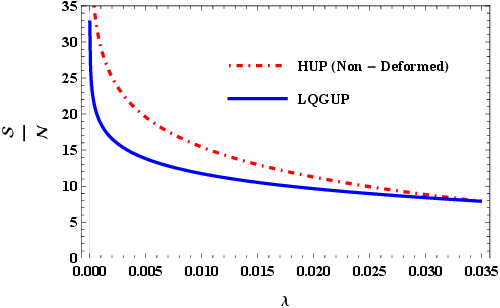}
    \caption{}
    \label{fig:sub2}
  \end{subfigure}
   \caption{The entropy of a 3D harmonic oscillator is shown in Figure (a) as a function of temperature. It is evident that, in the LQGUP model with minimal length and maximal momentum, the entropy increases at a slower rate than in the standard non-deformed case. This is due to the reduction in the number of accessible microstates in the high temperature regime, caused by the effects of quantum gravity. Figure (b) illustrates the entropy in relation to the thermal wavelength. It is observed that at extremely small wavelengths, the entropy demonstrates a slower growth rate compared to the classical (HUP) scenario.}
  \label{fig:4}
\end{figure}

According to Eq. $S=k_{B} \ln {\Omega}$, the entropy of each system is directly proportional to the number of accessible microstates. Therefore, it is reasonable to expect that the number of microstates in the GUP framework is lower than in the classical case. This can be observed in Figure (\ref{fig:4}), where the entropy of the system in the LQGUP model increases at a slower rate compared to its classical counterpart. This result was anticipated, as we have previously observed a reduction in the degrees of freedom in this model, from $6$ to $3$.

\subsection{Numerical Results}

The partition function obtained in equation (\ref{ho0}) can be solved exactly by numerical techniques using Mathematica. Subsequently, other  related thermodynamic variables can be determined exactly and their plots can be generated. In this subsection, in addition to numerical analysis encoded in the plots, we present some numerical results in the table (\ref{T:T1}). It will be observed that the approximate analytical results from the previous section closely align with the numeric findings. In the diagrams (\ref{fig:1},\ref{fig:2},\ref{fig:3},\ref{fig:4}), we have plotted the temperature-dependent rate of change for various thermodynamic parameters such as internal energy, number of degrees of freedom, specific heat and entropy. As can be seen from the diagrams and the table, the analytical values determined in the previous section are very close to the exact numerical result.

\begin{table}
  \centering
\begin{tabular}{|c|c|c|c|c|c|c|c|c|}
  \hline
  T & 20 & 40 & 60 & 80 & 100   \\\hline
  $\frac{U}{N}$ (analytical) & 61.6745 & 101.695 & 135.125 & 166.8 & 197.786   \\\hline
   $\frac{U}{N}$(numerical) & 60.0699 & 99.6916 & 133.094 & 164.78 & 195.778   \\\hline
  \hline
   T & 20 & 40 & 60 & 80 & 100\\\hline
  $\frac{2U}{NT}$(analytical) & 6.16745 & 5.08475 & 4.50417 & 4.17 & 3.95572   \\\hline \hline
 $\frac{2U}{NT}$(numerical) & 6.00699 & 4.98458 & 4.43648 & 4.11949 & 3.91557  \\\hline \hline
   T & 20 & 40 & 60 & 80 & 100\\\hline
  $\frac{C_V}{N}$ (analytical) & 2.42667 & 1.75936 & 1.61284 & 1.56211 & 1.53909  \\\hline
  $\frac{C_V}{N}$ (numerical) & 2.37137 & 1.75497 & 1.61298 & 1.56276 & 1.53973 \\\hline \hline
    T & 20 & 40 & 60 & 80 & 100 \\\hline
  $\frac{S}{N}$ (analytical) & 5.45347 & 6.8666 & 7.54631 & 8.00228 & 8.34809    \\\hline
  $\frac{S}{N}$ (numerical) & 5.28132 & 6.6787 & 7.3578 & 7.8139 & 8.15986    \\
  \hline
\end{tabular}
  \caption{This table contains the approximate and exact rates of change of internal energy, number of degrees of freedom, specific heat and entropy at different temperatures using both analytical and numerical methods for a system of a 3D harmonic oscillator in the LQGUP phase space model.}\label{NDF}
  \label{T:T1}
\end{table}

\section{The Case of 2D Harmonic Oscillator}

In this section we consider a 2D harmonic oscillator to see the reduction of the degrees of freedom in this system and whether the results are as expected or not.
The 2D LQGUP Liouville volume, $d^2\omega$, is obtained from equation (\ref{PB10}) as $d^2\omega = \frac{d^2{q}d^2{p}}{\big[1-\kappa p+\frac{7}{6}\kappa^2p^2\big]^{3}}$. Hence, for 2D harmonic oscillators in LQGUP phase space, the partition function of a single particle in the semi-classical regime can be derived as follows

\begin{eqnarray}\label{PB14-2}
\begin{gathered}
Z_1(T,V,\kappa ) = \frac{1}{h^2}\int_\Gamma d^2\omega\exp[-H/T]= \frac{1}{{{h^2}}}\int {{d^2}q \int{\frac{{{d^2}p}}
{\Big[1-\kappa p+\frac{7}{6}\kappa^2p^2\Big]^{3}}\exp \left( { - \frac{{H(q,p)}}{T}} \right)} }\,.
\end{gathered}
\end{eqnarray}
Expanding the deformed phase space volume yields
\begin{equation}\label{ho0.5-2}
  \frac{1}{\Big[1-\kappa p+\big(\frac{7}{6}\big)\kappa^2p^2\Big]^{3}} =1+3\kappa p+\frac{5 \kappa ^2 p^2}{2}+O\left(\kappa ^3\right).
\end{equation}
So, the partition function (\ref{PB14-2}) can be approximated as follows
\begin{equation}\label{ho1-2}
\begin{gathered}
 {\mathcal Z}_1\simeq \frac{4\pi^2}{h^2}\int_{0}^{\infty}\int_{0}^{\frac{1}{2\kappa}}
\exp\big[-\frac{p^2}{2mT}\big]\exp\big[-\frac{m{w^2}q^2}{2T}\big]qp \big(1+3\kappa p+\frac{5 \kappa ^2 p^2}{2}\big) dp dq
\end{gathered}
\end{equation}
Upon integration of the above relation, we obtain
\begin{equation}\label{ho1.5-2}
\begin{gathered}
{\mathcal Z}_1[T]=\Big(\frac{T}{w}\Big)^2\Bigg[1+5m T\kappa^2- \Big(\frac{25}{8}+5 m T \kappa ^2\Big)\exp \Big[{\frac{-1}{8m T\kappa^2 }}\Big]+\frac{3}{4}\sqrt{\pi}\sqrt{8m T\kappa^2} \text{erf}\Big(\frac{1}{ \sqrt{8m T\kappa^2}}\Big)\Bigg]\,.
\end{gathered}
\end{equation}
Expressing it in terms of the thermal de Broglie wavelength $\lambda$, we get
\begin{equation}\label{ho2-2}
\begin{gathered}
 {\mathcal Z}_1[\lambda]\approx \Big(\frac{2\pi \ell^2}{\lambda^2}\Big)^2\bigg[
[1+\frac{5{\lambda_{_{\rm Pl}}}^2}{8\lambda^2}- \Big(\frac{25}{8}+\frac{5{\lambda_{_{\rm Pl}}}^2}{8\lambda^2}\Big)\exp \Big[\frac{-\lambda^2}{\lambda_{_{\rm Pl}}^2}\Big]+\frac{3\sqrt{\pi}{\lambda_{_{\rm Pl}}}}{4\lambda} \text{erf}\Big[\frac{\lambda}{\lambda_{_{\rm Pl}}}\Big] \bigg].
 \end{gathered}
\end{equation}
Expanding the partition function (\ref{ho2-2}) at very high and low temperatures gives
\begin{equation}\label{ho3-2}
{\mathcal Z}_1[\lambda]=\left\{
  \begin{array}{ll}
    \begin{gathered}(4{\pi}^2)\frac{{\ell}^4}{\lambda^4}=\Big(\frac{T}{ \omega}\Big)^2\end{gathered} & \hspace{.5cm}\lambda
    \gg\lambda_{_{\rm Pl}},\\\\
    \begin{gathered}\Big(\frac{37\pi^2}{4}\Big)\frac{\ell^4}{\lambda^2\lambda_{_{\rm Pl}}^2}\end{gathered} &
    \hspace{.5cm}\lambda\sim\lambda_{_{\rm Pl}}\,,
  \end{array}
\right.
\end{equation}
It is evident from Eq.~(\ref{ho3-2}) that two degrees of freedom will freeze ($\lambda^4\rightarrow\lambda^2$) at the Planck length scale $\lambda\sim\lambda_{_{\rm Pl}}$ in the LQGUP framework, and the partition function behaves as if it were effectively a partition function of a $1$D harmonic oscillator.\\
The internal energy of the system can be modified as follows
\begin{eqnarray}\label{ho5-2}
U=NT\left(
\frac{\begin{gathered}-25-\frac{{61\lambda_{_{\rm Pl}}}^2}{\lambda^2}-\frac{15{\lambda_{_{\rm Pl}}}^4} {\lambda^4} + \frac {2 {\lambda_{_{\rm Pl}}}^2}{\lambda^2}\exp\Big[\frac{\lambda^2}{\lambda_{_{\rm Pl}}^2}\Big]\bigg(8+\frac{15{\lambda_{_{\rm Pl}}}^2}{2\lambda^2}+\frac{15\sqrt{\pi}{\lambda_{_{\rm Pl}}}}{2\lambda}\text{erf}\Big[\frac{\lambda}{\lambda_{_{\rm Pl}}}\Big]\bigg)\end{gathered}}
{\begin{gathered}\frac{{\lambda_{_{\rm Pl}}}^2}{\lambda^2}\bigg(-5(5+\frac{{\lambda_{_{\rm Pl}}}^2}{\lambda^2})+ 4\exp\Big[\frac{\lambda^2}{\lambda_{_{\rm Pl}}^2}\Big]\Big(2+\frac{5{\lambda_{_{\rm Pl}}}^2}{4\lambda^2}+\frac{3\sqrt{\pi}{\lambda_{_{\rm Pl}}}}{2\lambda}\text{erf}\Big[\frac{\lambda}{\lambda_{_{\rm Pl}}}\Big]\Big)\bigg)\end{gathered}}\right).
\end{eqnarray}
Expanding the relation (\ref{ho5-2}) for both of the high and low temperature regimes, we have.
\begin{equation}\label{ho6-2}
U=
\left\{
  \begin{array}{ll}
    2\,NT & \hspace{.5cm}\lambda
    \gg\lambda_{_{\rm Pl}},\\\\
   \,NT &
    \hspace{.5cm}\lambda\sim\lambda_{_{\rm Pl}}\,.
  \end{array}
\right.
\end{equation}
Finally, expanding the specific heat which is defined through the relation (\ref{ho6-2}) is as follows
\begin{equation}\label{ho8-2}
C_{_V}=
\left\{
  \begin{array}{ll}
    2\,N & \hspace{.5cm}\lambda
    \gg\lambda_{_{\rm Pl}},\\\\
   \,N &
    \hspace{.5cm}\lambda\sim\lambda_{_{\rm Pl}}.
  \end{array}
\right.
\end{equation}

These results for a system of 2-dimensional harmonic oscillator are in agreement with the general prescription of the reduction of the degrees of freedom as expected.

\begin{figure}[ht]
  \centering
  \begin{subfigure}[b]{0.35\textwidth}
    \includegraphics[width=3.3in]{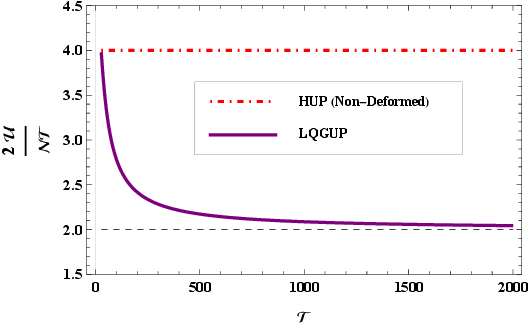}
    \caption{}
    \label{fig:sub1}
  \end{subfigure}
  \hfill
  \begin{subfigure}[b]{0.48\textwidth}
    \includegraphics[width=3.3in]{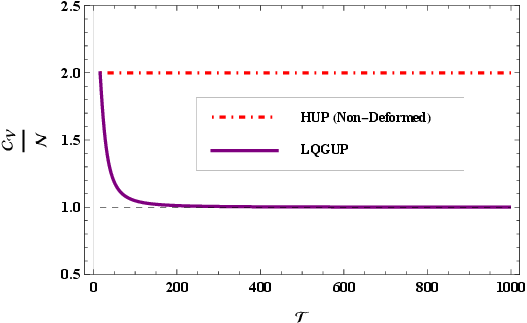}
    \caption{}
    \label{fig:sub2}
  \end{subfigure}
  \caption{The number of degrees of freedom of a 2D harmonic oscillator is plotted in Figure (a) as a function of temperature. It is observed that in the LQGUP model, this value is temperature-dependent and decreases from $4$ to $2$ at extremely high temperatures for a 2D harmonic oscillator system.  Figure (b) shows the specific heat of a 2D harmonic oscillator as a function of temperature. The system's specific heat is temperature-dependent for the LQGUP model and asymptotically approaches $1$ at the very high temperature regime. It is also clear from the figure that the specific heat is bounded as $1 \leq\frac{C_{_V}}{N}\leq 2$ in the LQGUP framework.}
  \label{fig:2-2}
\end{figure}

\section{Summary and Conclusions}

In this paper, we have investigated the thermostatistics of a harmonic oscillator system in the Linear Quadratic GUP phase space model. Our aim was to gain further insight into the topic of the reduction of the degrees of freedom in high energy regimes. The calculations have been based on the existence of an ultraviolet cutoff (or minimal measurable length and maximal measurable momentum). Referring to the equipartition theorem of energy, we have clearly shown that as the temperature increases towards the Planck temperature in the mentioned framework, the three degrees of freedom freeze out. Such freezing (reduction) of the degrees of freedom has been observed for the Snyder Phase Space framework in a previous work by our research group~\cite{NHG15}. The most important observation here is the fact that, while in the Snyder phase space scheme, the reduction of the number of degrees of freedom at high temperatures is $2$-folds, in the GUP framework, this reduction is $3$-folds.\\
In this direction, we note the following important points: \\
$\bullet$ First, we note that the reduction of the degrees of freedom is essentially continuous, which means the possibility of having a non-integer degrees of freedom (see also figures \ref{fig:2}). This may signal a fractal space dimension in the presence of the phenomenological quantum gravity effect encoded in the presence of a minimal measurable length.\\
$\bullet$ Second, we considered the classical equipartition theorem for our treatment. It is in fact necessary to consider a modified equipartition theorem and even a Planck scale thermodynamics that may be in dramatic digression with the standard thermodynamics. But these are the issues of the yet-unsolved quantum gravity problem.\\
 In the absence of a well-formulated quantum gravity theory, such phenomenological-based studies may shed light on the issue of the final quantum gravity proposal. So, we emphasize that in general, the deformed spaces in different phenomenological approaches to quantum gravity proposal all address a reduction of the degrees of freedom in high temperature regimes from a thermodynamic point of view. All of these theories coincide with their classical counterparts at low temperatures, as required. The possibility of realizing the fractal degrees of freedom in these setups is also an important outcome of this study.\\
 The behavior shown in the diagrams arises in the field of quantum gravity, where the effects of the generalized uncertainty principle (GUP) lead to changes in the structure of phase space. The study of the LQGUP deformation scenario reveals a remarkable aspect: a significant deviation from classical behavior in the number of microstates at high energies. This deviation, characterized by the blocking of degrees of freedom, underlines a complex interplay between the basic principles of quantum mechanics and gravity. The shift towards a low-dimensional phase space at high energies emphasizes the intricate correlation between the dynamics of microstates and the inherent quantum essence of spacetime. These insights provide valuable perspectives on the nature of spacetime in a realm where the influences of quantum gravity are significant and provide a deeper understanding of the fundamental structure of the universe.\\
 Finally we have checked the case with 2-dimensional harmonic oscillator to see the results which are in agreement with the general prescription: a reduction of the degrees of freedom from $4$ to $2$, as expected.\\

 {\bf Acknowledgement}\\
 We appreciate the referee for his/her insightful and constructive comments. \\

\end{document}